\documentclass[sigconf]{acmart}
\usepackage{booktabs}
\usepackage{multirow}
\usepackage{multicol}
\usepackage{balance}
\AtBeginDocument{%
  \providecommand\BibTeX{{%
    \normalfont B\kern-0.5em{\scshape i\kern-0.25em b}\kern-0.8em\TeX}}}
\copyrightyear{2022}
\acmYear{2022}
\setcopyright{acmcopyright}
\acmConference[SIGIR '22]{Proceedings of the 45th International ACM SIGIR Conference on Research and Development in Information Retrieval}{July 11--15, 2022}{Madrid, Spain}
\acmBooktitle{Proceedings of the 45th International ACM SIGIR Conference on Research and Development in Information Retrieval (SIGIR '22), July 11--15, 2022, Madrid, Spain}
\acmPrice{15.00}
\acmDOI{10.1145/3477495.3531932}
\acmISBN{978-1-4503-8732-3/22/07}

\begin{document}
\fancyhead{}

\title{Adaptable Text Matching via Meta-Weight Regulator}

\author{Bo Zhang}
\email{bo.zhang@bit.edu.cn}
\affiliation{
  \institution{Beijing Institute of Technology}
  \city{Beijing}
  \state{Beijing}
  \country{China}
}
\author{Chen Zhang}
\email{czhang@bit.edu.cn}
\affiliation{
  \institution{Beijing Institute of Technology}
  \city{Beijing}
  \state{Beijing}
  \country{China}
}
\author{Fang Ma}
\email{mfang@bit.edu.cn}
\affiliation{
  \institution{Beijing Institute of Technology}
  \city{Beijing}
  \state{Beijing}
  \country{China}
}
\author{Dawei Song}
\email{dwsong@bit.edu.cn}
\authornote{Corresponding author. Also with School of Computing and Communications, The Open University, United Kingdom}
\affiliation{
  \institution{Beijing Institute of Technology}
  \city{Beijing}
  \state{Beijing}
  \country{China}
}
\renewcommand{\shortauthors}{Zhang et al.}
\begin{abstract}
    Neural text matching models have been used in a range of applications such as question answering and natural language inference, and have yielded a good performance. However, these neural models are of a limited adaptability, resulting in a decline in performance when encountering test examples from a different dataset or even a different task. The adaptability is particularly important in the few-shot setting: in many cases, there is only a limited amount of labeled data available for a target dataset or task, while we may have access to a richly labeled source dataset or task. However, adapting a model trained on the abundant source data to a few-shot target dataset or task is challenging. To tackle this challenge, we propose a Meta-Weight Regulator (MWR), which is a meta-learning approach that learns to assign weights to the source examples based on their relevance to the target loss. Specifically, MWR first trains the model on the uniformly weighted source examples, and measures the efficacy of the model on the target examples via a loss function. By iteratively performing a (meta) gradient descent, high-order gradients are propagated to the source examples. These gradients are then used to update the weights of source examples, in a way that is relevant to the target performance. As MWR is model-agnostic, it can be applied to any backbone neural model. Extensive experiments are conducted with various backbone text matching models, on four widely used datasets and two tasks. The results demonstrate that our proposed approach significantly outperforms a number of existing adaptation methods and effectively improves the cross-dataset and cross-task adaptability of the neural text matching models in the few-shot setting.
\end{abstract}


\maketitle

\section{Introduction}
    Text matching aims to identify the relationship between two text fragments. It has been a key research problem in natural language processing (NLP) and information retrieval (IR). A broad range of tasks can be viewed as specific forms of text matching, such as question answering \citep{wang2007jeopardy, yang2015wikiqa} and natural language inference \citep{bowman2015large}.

    With the rapid development of deep learning techniques, many neural text matching models have been proposed in recent years and have achieved an impressive performance on various tasks, due to their capacity of learning text representations and modeling interactions of text pairs. However, deep learning-based methods tend to be data-hungry and have a strong dependency on the scale of labeled training data. In many cases, however, there is only a limited amount of labeled data available for a target data distribution (in term of \textit{dataset} or \textit{task}), so that a poor performance could result. On the other hand, we may have access to a richly labeled source data distribution, on which we can train a model and then adapt it to the target data. This scenario is known as the \textit{few-shot setting}, in which the adaptability of the model is particularly important. In general, the challenge lies in  how to make a model learn effectively on a small amount of labeled examples in the target data distribution, with the aid of the richly labeled source data, where the two distributions may be largely different. 
    
    Some recent attempts have been made to address the problem \citep{li2016revisiting, zhao2020domain, balaji2020robust, zhang2020domain, yang2020fda, asghar2018progressive, cohen2018cross}. 
    A straightforward way is to merge the source and target data for training. Still, previous related work \citep{koehn2017six} has shown that the source examples related to the target data are beneficial to, while the examples not associated with the target data may impair the model's performance. Further, a domain confusion method based on adversarial learning is proposed for domain adaptation \citep{cohen2018cross}. The method learns a domain classifier while training, and forces the text matching model to perform the gradient descent to confuse the source and target domains by adversarial learning. Since a domain classifier needs to be learned, a natural precondition for this approach is that the amounts of data in the source and target domains should be as comparable as possible, i.e., complying  with an inter-domain class balance. This contradicts our application scenarios, where it is assumed not practical to collect and label the target data to a scale comparable to the source data. Apart from that, a reinforcement learning-based data selector is presented for cross-domain transfer learning \citep{qu2019learning}. Particularly, this method first selects high-quality data on the richly-labeled source domain and then performs transfer learning on the selected target data. Although this method is suitable for limited target labeled data, the model's training on the source data and target data is not performed simultaneously, which increases the training time. More importantly, the data selection is only made at the data level rather than the gradient level, which could become meaningless if the distribution gap between the source and target data is large.
    
    Most of the adaptation methods in previous works are claimed to have experimented in cross-domain settings. Although data distribution across the source and target domains differ, the differences are not very large. In general, the source and target domains may be data on different topics of the same dataset or different datasets of the same task. Nonetheless, it is still challenging to obtain datasets of the same task in practical application scenarios. Therefore, to cope with enormous data distribution gaps, we need to eliminate cross-domain limitations and generalize the adaptation method to more general adaptation scenarios, such as cross-task adaptation. This paper refers to the approaches in general scenarios as adaptation methods on different data distributions to distinguish them from cross-domain adaptation methods.

    In this paper, we present a Meta-Weight Regulator to enhance the adaptability of text matching models in the few-shot setting. The core idea is to relevantly weight the source examples along the gradient descent directions that minimize the loss of the target examples. In contrast to traditional training, the Meta-Weight Regulator adopts the meta-learning to achieve the goal of weighting source examples. Specifically, our regulator first assigns an initialization weight to each source data and takes the high-order derivatives of the source weights derived from the loss of the target examples. Then, it regulates the source weights through a meta-gradient descent according to the higher-order derivatives and computes the weighted source loss. Compared to the previous methods, our method imposes no requirement for the scale of the target data, as there is no need to train a domain classifier. This method can also be regarded as a data selection process guided by the gradient descent at the gradient level. Since the weights of the source samples are regulated according to the target loss during the training process, it ensures that the text matching model is trained on the source data and the target data simultaneously, which can also improve the training efficiency. For simplicity and efficiency of the method, we borrow the idea of higher-order derivatives from Model-Agnostic Meta-Learning \citep{finn2017model}. Accordingly, the Meta-Weight Regulator is model-agnostic so that we can use it on top of an existing neural model to improve their adaptability and does not require any extra hyperparameter tuning.
    
    We conduct extensive experiments on two text matching tasks, namely Natural Language Inference and Question Answering, using four widely used datasets, i.e., SNLI \citep{bowman2015large}, SCITail \citep{khot2018scitail}, TrecQA \citep{wang2007jeopardy} and WikiQA \citep{yang2015wikiqa}, where they are all used as the sources and targets respectively to simulate the few-shot and many-shot setting. The Meta-Weight Regulator outperforms various state-of-the-art adaptation methods in term of cross-dataset and cross-task adaptability, and brings significant improvements over a wide range of backbone neural text matching models on the target test sets in almost all experiments. 
    
    To the best of our knowledge, the proposed regulator is the first approach to perform cross-task adaptation on deep text matching models in the few-shot setting. Our method can effectively regulate text matching models for adaptation if the data distribution and label categories between the source and the target are different.

\section{Related Work}
    This section presents an overview of related work in deep text matching models, few-shot adaptation methods, and methods for weighting examples.

    \subsection{Deep Text Matching}
        Over the past few years, numerous deep learning based text matching models have been proposed. They can be roughly divided into two categories: representation-based \citep{feng2015applying, mueller2016siamese, nie2017shortcut} and interaction-based \citep{pang2016text, chen2016enhanced, devlin2018bert}. 
        
        The representation-based models first leverage neural networks to obtain the representation of each text separately and then predict the relationship between a pair of text by calculating the distance between their representations in a latent semantic space. Typical examples of this type of models include SiameseLSTM \citep{mueller2016siamese}, SSE \citep{nie2017shortcut} and Sentence-BERT \citep{reimers2019sentence}. The idea of interaction-based models is to model the interactions between two paragraphs of text from different levels of granularity (e.g. word level, phrase level) using various attention methods, and aggregate them to form a feature representation of the overall interaction that is then fed into deep neural networks to further extract features and make prediction. Typical models that follow this idea include MatchPyramid \citep{pang2016text}, ESIM \citep{chen2016enhanced}, and BERT \citep{devlin2018bert}. 
        
        In this work, we mainly use BERT, a state-of-the-art neural model of text matching, as the backbone model, and perform the proposed Meta-Weight Regulator on top of it, to compare against other adaptation techniques in the few-shot setting. Meanwhile, we also test the proposed method on top of various other representative backbone models.

    \subsection{Few-Shot Adaptation Methods}
        Few-shot learning, which aims at learning with a limited number of labeled examples, is a meaningful way to bridge the gap between AI and humans and relieve the burden of collecting and annotating large-scale labeled data. This is especially the case for text matching tasks, as the volume of sentence-pair data is considerably large and the manual determination of the semantic relationships for each pair of text is costly. It would be more efficient to solve the current few-shot learning problem with the help of a  trained model on another label-rich task or dataset than spending a lot of time and labor costs in constructing a new dataset for the current task. In this way, the problem of few-shot learning is transformed into a specific form of adaptation methods, which aim to eliminate the discrepancy between the source and target data distributions. 
        
        One stream of adaptation methods is to learn domain-confusion representations with domain classifiers by adversarial training \citep{ganin2016domain, cohen2018cross, qu2019adversarial, xue2020improving}. Nevertheless, this type of methods may cause the problem of unbalanced class bias, because the amount of source data is far more than the target data in the few-shot setting. Another type of methods is to learn a source data selector by reinforcement learning to facilitate the transfer learning of models from the source to the target data \citep{qu2019learning}. Although this method does not require a large scale target data, the model needs to be trained on the source and the target data sequentially for transfer learning, which increases the training time. At the same time, data selection is carried out at the data level instead of the gradient level. When the distributions of source and target data are quite different, such as cross tasks, improving the model's adaptability would be particularly challenging.
        
        This paper proposes the adaptation method Meta-Weight Regulator to help text matching models improve adaptation in a few-shot learning setting. Unlike data selection only at the data level, our approach conducts data selection at the gradient level through a meta-learning paradigm and trains the model more efficiently while achieving better adaptive performance.

\begin{figure*}
    \centering
    \includegraphics[width=0.9968\textwidth, height=0.3626\textwidth]{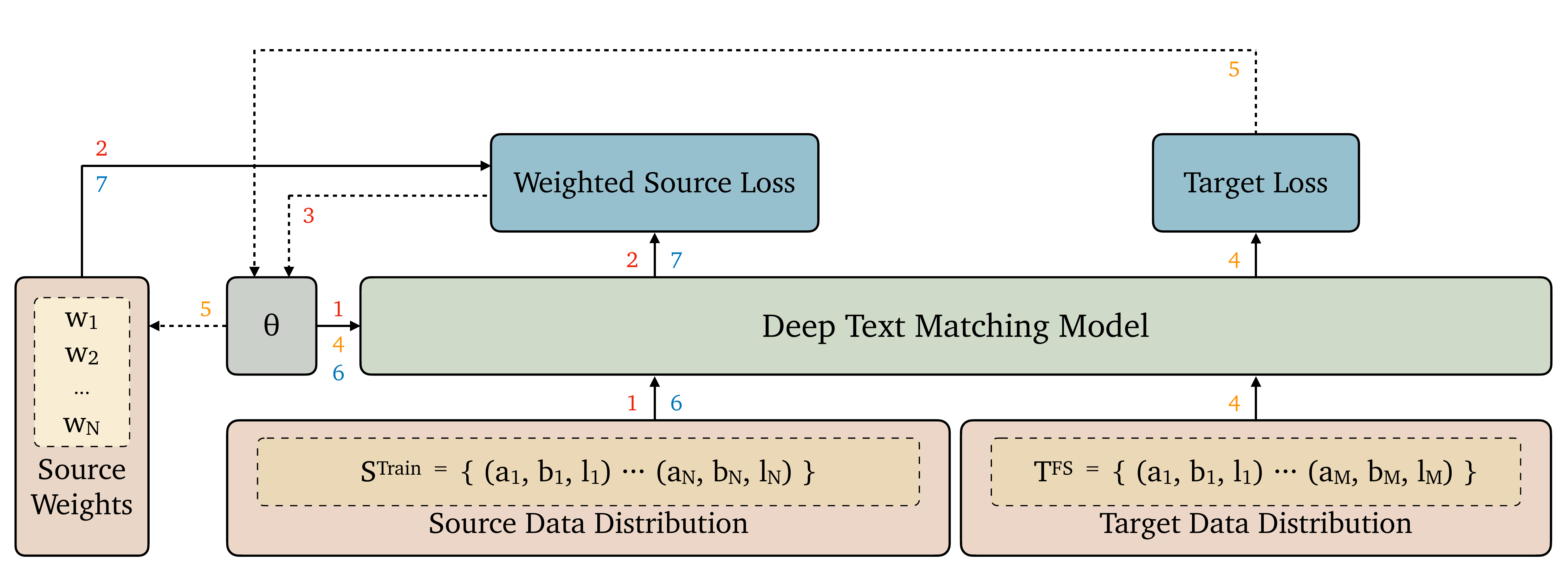}
    \caption{The algorithm of Meta-Weight Regulator. The solid arrows represent the forward propagation process, and the dashed arrows represent the backpropagation process. $\theta$ indicates the learnable parameters of the deep text matching model. The numbers marked on the arrows represent the indices of their corresponding equations in Section \ref{section: algorithm}. The red, orange and blue numbers correspond to the equations described in Sections \ref{section: algorithm_1}, \ref{section: algorithm_2} and \ref{section: algorithm_3}, separately.}
    \label{figure:1}
\end{figure*}

    \subsection{Weighting Examples}
        The idea of weighting training examples has been studied extensively in the literature. Focal loss adds a soft weighting scheme that emphasizes harder examples \citep{lin2017focal}. Likewise, boosting algorithms such as AdaBoost select harder examples to train subsequent classifiers \citep{freund1997decision}. Hard example mining downsamples the majority class and utilizes the most difficult examples \citep{malisiewicz2011ensemble}. On the other hand, hard examples are not always preferred in the presence of outliers and noisy processes. Robust loss estimators typically downgrade examples with a high loss. In self-paced learning, example weights are obtained by optimizing a weighted training loss, thus promoting the easier examples to be learned first \citep{kumar2010self}. Importance sampling is a classic method in statistics that assigns weights to samples to match one distribution with another \citep{kahn1953methods}. Weighting examples is also applied to curriculum learning and regularization of noisy and corrupted labels \citep{bengio2009curriculum, jiangmentornet, ren2018learning}.
        
        The above works on weighting samples involve data distribution matching, difficult sample mining, increasing or decreasing sample weights according to loss, and weighted loss calculation. The Meta-Weight Regulator in this paper employs the paradigm of meta-learning to bring these aspects  together to address the few-shot adaptation problem of text matching. Our approach narrows the distribution gap between source and target data by regulating the weights assigned to source examples based on the target training loss.

\section{Methodology}
    In this section, we present the proposed Meta-Weight Regulator for the few-shot text matching in the cross-task and cross-dataset adaptation setting. We first describe the application scenarios and problem setting, followed by the Meta-Weight Regulator which narrows the distribution gap between the source and target data by regulating the weights assigned to source examples based on the target training loss.

    \subsection{Application Scenarios}
        Deep text matching models have been widely used in many application tasks, for example, question answering and natural language inference, which will be introduced in more detail in Section \ref{section: data}. The models used for these applications are commonly trained on the application-related data. When faced with a new application task, new application-related data needs to be collected and labeled, due to the data-hungry nature of deep learning models and their dependency on the source data distribution. However, collecting and labeling such large scale data requires enormous efforts and should be avoided if possible. In this scenario, it is of theoretical and practical significance to adapt a model trained on a source dataset or task to the new application by using a small amount of labeled new application-related data. Our Meta-Weight Regulator aims to address this problem. In this paper, we attempt to use a Meta-Weight Regulator and a small number of labeled target samples, so that a deep text matching model trained on the source data can adapt to the target effectively. 
        
    \subsection{Problem Setting}
        This work is aimed to enhance the adaptability of deep text matching models in the few-shot learning setting. We assume that there exist a set of source samples and a set of target samples in the training phase, where the distributions of the two sets are inconsistent. In the former, there are a large number of labeled samples available that are sufficient for regular deep learning training. Accordingly, the train set of the source distribution is denoted as ${S^{Train}}$. In the latter, there are only a small amount of labeled samples, which form a few-shot target dataset, denoted as ${T^{FS}}$. The goal of our work is to regulate the model trained on ${S^{Train}}$ by using ${T^{FS}}$, such that the model can adapt well to the target distribution after training. For instance, we can train a model on numerous Natural Language Inference datasets \cite{bowman2015large, khot2018scitail} and leverage a small amount of Question Answering data \citep{yang2015wikiqa, wang2007jeopardy} to regulate the training process, then evaluate the model's adaptability from natural language inference (source) to question answering (target).
        
        Formally, each text matching training sample is denoted as $\left(a, b, l\right)$, where $a$ and $b$ denote a text pair and $l$ denote the label for the pair. Hence we can respectively formalize ${S^{Train}}$ as $\left\{ \left( a_{i}^{s}, b_{i}^{s}, l_{i}^{s} \right), 1 \leq i \leq N \right\}$ and ${T^{FS}}$ as $\left\{ \left( a_{j}^{t}, b_{j}^{t}, l_{j}^{t} \right), 1 \leq j \leq M \right\}$, and suppose $M \ll N$. Any deep text matching model used as the backbone of our regulator is noted as $F \left( a, b, \theta \right)$, where $\theta$ represents the model parameters. The loss function is denoted as $Loss \left( y, l \right)$ where $y = F \left( a, b, \theta \right)$. Hereafter, we will continue with these formal expressions for simplicity, and use the superscripts $s$ and $t$ correspondingly to denote the related source and target datasets or tasks, while the subscripts $i$ and $j$  indicate the $i^{th}$ and $j^{th}$ data examples respectively.

    \subsection{Meta-Weight Regulator}
        \label{section: algorithm}
        Our proposed method, called Meta-Weight Regulator, is a meta-learning-based algorithm that regulates the weight of each sample when computing the loss function. The regulation process is performed according to the gradient descent direction during the training phase. Specifically, the Meta-Weight Regulator contrasts the gradient descent direction of the text matching model on ${S^{Train}}$ and ${T^{FS}}$ to regulate the weight of the source sample in the loss function, and finally calculate the weighted source loss. During the regulating process, the source weight is increased when the gradient descent directions of the source and target samples are similar, and vice versa, so that the model trained on the source data learns to minimize the target loss. From another perspective, this weight regulating process can be seen as the process of using ${T^{FS}}$ to perform soft data selection on ${S^{Train}}$, which is performed at the gradient level. We adopt the meta-learning paradigm to establish a computational graph connection between source weights and target losses for gradient comparison and source weight regulation. At the same time, the model-agnostic meta learning based algorithm ensures that our approach does not introduce additional hyperparameters. As illustrated in Figure \ref{figure:1}, the Meta-Weight Regulator first assigns a weight to each source sample and associates the source weight with the target loss in the computation graph. Next, the source weights are updated by gradient descent via the second derivative of the target loss. Finally, the model is trained on the weighted source samples, and the model parameters are updated according to the weighted source loss. In the rest of the section, the Meta-Weight Regulator will be described in more detail step by step.

    \subsubsection{\textbf{Establishing A Connection in the Computational Graph}}
        \label{section: algorithm_1}
        First, our method assigns an initialization weight to each source sample. Considering that the difference between the source and target data distribution is vast, we initialize each source weight as 0 to prevent early update of the model parameters on the source distribution from falling into local values unfavorable to the target performance. Through subsequent calculations, the weights of the source data that are beneficial to the model's performance on the target distribution are increased.
        
        To establish a computational graph relationship between the source weights and the target distribution loss during training, we first establish a relationship of the source weight with the model parameters. To do so, we need to update the text matching model parameters on the weighted source data. Specifically, the source weights are applied to compute the model's weighted source loss:
        \begin{equation}
            y^s_i = TMM^s\left(a^s_i,b^s_i,\theta\right)
            \label{equation:1}
        \end{equation}
        \begin{equation}
            Loss^s\left(y^s ,l^s\right) = \sum\limits_{i=1}^N w^s_i Cost^s\left(y^s_i, l^s_i\right)
            \label{equation:2}
        \end{equation}
        where $i \in \{ 1, 2, ..., N \}$, $a^s_i, b^s_i \in S^{Train}$, $TMM^s$ indicates the text matching model for forwarding propagation on $S^{Train}$, $\theta$ denotes the model parameters, and $Cost$ is the cost function of the text matching model.
        
        Then the model parameters are updated by gradient descent. The updated model parameters can be formalized as dependent variables with source weights as independent variables:
        \begin{equation}
            \begin{aligned}
            \widetilde{\theta} & = \theta - \alpha \cdot \frac{\partial Loss^s\left(y^s ,l^s\right)}{\partial \theta} \\
            & = \theta - \alpha \cdot \frac{\partial \sum\limits_{i=1}^N w^s_i Cost^s\left(y^s_i, l^s_i\right)}{\partial \theta}
            \end{aligned}
            \label{equation:3}
        \end{equation}
        where $\widetilde{\theta}$ represents the updated model parameters, and $\alpha$ indicates the learning rate.
        
\begin{figure}
            \centering
            \includegraphics[width=0.48633\textwidth, height=0.11339\textwidth]{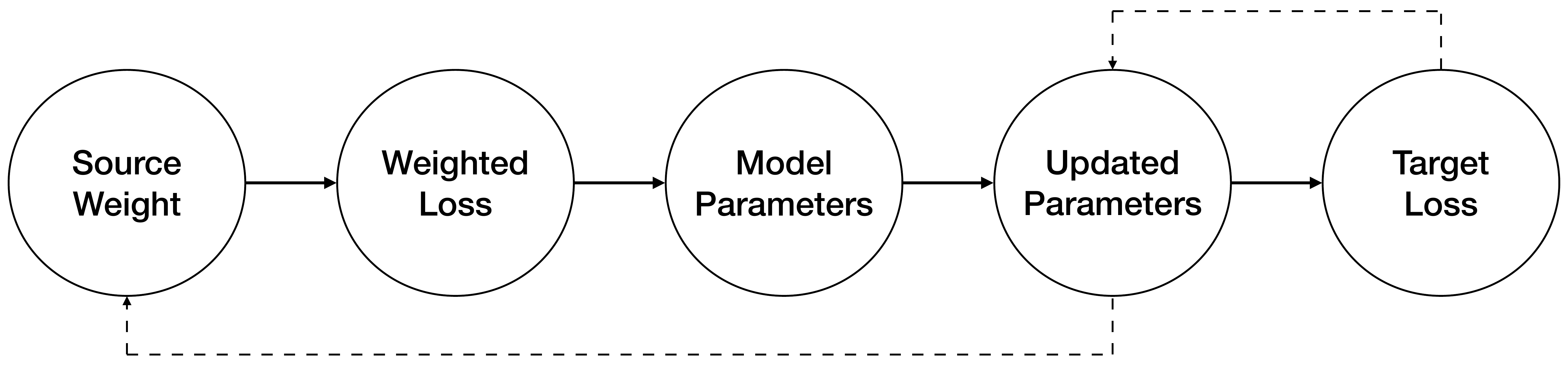}
            \caption{The Path of Computational Graph. The solid and dotted arrows represent the forward and backward propagation paths.}
            \label{figure:2}
        \end{figure}

        Since both the initialized source weight and the weighted source loss are 0, the updated model parameters do not change numerically, but there is a path from the source weights to the updated model parameters in the computation graph, as shown in Figure \ref{figure:2}.
        
    \subsubsection{\textbf{Regulating Weights Through Meta-Gradient Descent}}
        \label{section: algorithm_2}
        After establishing a computational graph connection between source weights and model parameters, the Meta-Weight Regulator regulates the source weights by comparing the source and target gradient descent directions. Inspired by MAML \citep{finn2017model}, we adopt the second derivative to contrast the gradient directions and update the source weights.
        
        Firstly, we train the text matching model on the target few-shot set and calculate the training loss:
        \begin{equation}
            \begin{aligned}
            Loss^t\left(y^t,l^t\right) & = \sum\limits_{j=1}^M Cost^t\left(y^t_j, l^t_j\right) \\ 
            & = \sum\limits_{j=1}^M Cost^t\left(TMM^t(a^t_j, b^t_j, \widetilde{\theta}), l^t_j\right)
            \end{aligned}
            \label{equation:4}
        \end{equation}
        where $a^t_j, b^t_j \in T^{FS}$, $TMM^t$ indicates the text matching model for forwarding propagation on $T^{FS}$.

        Since the source weights are linked to the updated parameters in the computation graph, when the Meta-Weight Regulator performs model gradient descent on the target data, the derivative of the target loss to the model parameters can naturally flow further to the source weights. This process is illustrated in Figure \ref{figure:2}. Therefore, the Meta-Weight Regulator can regulate the source weights by gradient descent according to the second derivative of the target loss:
        \begin{equation}
            \begin{aligned}
            \widetilde{w^s} & = w^s - \alpha \cdot \frac{\partial^2 Loss^t(y^t, l^t)}{\partial^2 w^s}\\
            & = w^s - \alpha \cdot \frac{\partial (\frac{\partial Loss^t(y^t, l^t)}{\partial \widetilde{\theta}} \cdot \frac{\partial \widetilde{\theta}}{\partial w^s})}{\partial w^s}
            \end{aligned}
            \label{equation:5}
        \end{equation}
        where $\alpha$ denotes the learning rate equaled with Eq.\ref{equation:3}.

        The above regulating method, denoted as meta-gradient descent, comes from meta-learning, which allows comparing the gradient descent directions of model parameters on different data distributions. The Meta-Weight Regulator adjusts the source sample weights, specifically by increasing the source weights, to minimize the target loss. In other words, the source samples that make the model parameters gradient descent in a similar direction to the target data's will be weighted higher, and vice versa. After the weight regulation, we integrate the updated source weights into the loss computation for regular model training on the source data.

    \subsubsection{\textbf{Training Text Matching Models on Weighted Samples.}}
        \label{section: algorithm_3}
        Lastly, we perform formal text matching model training on the source data and weight the source loss with the regulated weights to improve the model's adaptability:
        \begin{equation}
            y^s_i = TMM^s\left(a^s_i,b^s_i,\theta\right)
            \label{equation:6}
        \end{equation}
        \begin{equation}
            Loss^s(y, l) = \sum\limits_{i=1}^N \widetilde{w^s_i} \cdot Cost^s\left(y^s_i,l^s_i\right)
            \label{equation:7}
        \end{equation}
        where $i \in \{ 1, 2, ..., N \}$.

        In summary, the Meta-Weight Regulator operates based on the gradient descent algorithm, and the adjusted source weights are only used for the calculation of loss. Therefore it is suitable for any neural text matching models where the parameters are updated based on the gradient descent algorithm or its variants \citep{cherry1998sgd, kingma2014adam, loshchilov2017decoupled}. In other words, it is model-agnostic and complementary with any existing deep text matching model. Moreover, since the proposed regulator is implemented in a meta-learning paradigm that is similar to Model-Agnostic Meta-Learning \citep{finn2017model}, it can automatically regulate the model parameters and avoid introducing additional hyper-parameters that need to be manually tuned.

\begin{table}
	\centering
	\begin{tabular}{l|cccc}  
	    \toprule
        \textbf{Dataset} & \textbf{Tasks} & \textbf{Labels} & \textbf{Train} & \textbf{Test} \\
	    \midrule
	    WikiQA  & QA  & 2 & 20,360  & 6,165  \\
	    TrecQA  & QA  & 2 & 44,647  & 1,517  \\
	    SCITail & NLI & 2 & 23,596  & 2,126  \\
	    SNLI    & NLI & 3 & 549,367 & 10,000 \\
	    \bottomrule
	\end{tabular}
	\caption{Raw statistics of the four datasets in our experiments.}
	\label{table:1}
\end{table}

\begin{table}
	\centering
	\begin{tabular}{l|cccc}  
	    \toprule
        \textbf{Dataset} & \textbf{Tasks} & \textbf{Labels} & \textbf{Train} & \textbf{Test} \\
	    \midrule
	    WikiQA  & QA  & 2 & 2,078   & 582  \\
	    TrecQA  & QA  & 2 & 9,664   & 568  \\
	    SCITail & NLI & 2 & 16,944  & 1,684 \\
	    SNLI    & NLI & 2 & 366,768 & 6,666 \\
	    \bottomrule
	\end{tabular}
	\caption{Processed Statistics of the four datasets used in our experiments.}
	\label{table:2}
\end{table}

\begin{table*}[ht!]
	\centering
	\begin{tabular}{clcccc}  
	    \toprule
	    \multirow{2}{*}{\textbf{Target Size}} & \multirow{2}{*}{\textbf{Methods}} & \multicolumn{2}{c}{\textbf{Natural Language Inference}} & \multicolumn{2}{c}{\textbf{Question Answering}} \\ \cmidrule(r){3-4} \cmidrule(r){5-6}
        & & \textbf{SciTail} $\rightarrow$ \textbf{SNLI} & \textbf{SNLI} $\rightarrow$ \textbf{SciTail} & \textbf{WikiQA} $\rightarrow$ \textbf{TrecQA} & \textbf{TrecQA} $\rightarrow$ \textbf{WikiQA}\\ 
	    \midrule
    	\multirow{6}{*}{10-shot} & Backbone & 0.5391 & 0.5502 & 0.5307 & 0.5197 \\
    	& Fine Tuning & 0.5390 & 0.5505 & 0.5379 & 0.5306 \\
    	& Data Merging & 0.5392 & 0.5499 & 0.5304 & 0.5299 \\
    	& Reinforced Transfer Learning\citep{qu2019learning} & 0.5675 & 0.5796 & 0.5603 & 0.5538 \\
    	& Adversarial Domain Confusion\citep{cohen2018cross} & 0.5634 & 0.5426 & 0.5406 & 0.5495 \\
    	& Meta-Weight Regulator & $\textbf{0.5769}^\dagger$ & $\textbf{0.5887}^\dagger$ & $\textbf{0.5768}^\dagger$ & $\textbf{0.5796}^\dagger$ \\
    	\cline{0-5}
    	\multirow{6}{*}{50-shot} & Backbone & 0.5456 & 0.5551 & 0.5376 & 0.5285 \\
    	& Fine Tuning & 0.5463 & 0.5579 & 0.5380 & 0.5396 \\
    	& Data Merging & 0.5406 & 0.5288 & 0.5377 & 0.5297 \\
    	& Reinforced Transfer Learning\citep{qu2019learning} & 0.5725 & 0.5801 & 0.5632 & 0.5673 \\
    	& Adversarial Domain Confusion\citep{cohen2018cross} & 0.5642 & 0.5536 & 0.5411 & 0.5502 \\
    	& Meta-Weight Regulator & $\textbf{0.5865}^\dagger$ & $\textbf{0.5927}^\dagger$ & $\textbf{0.5858}^\dagger$ & $\textbf{0.5833}^\dagger$ \\
    	\cline{0-5}
    	\multirow{6}{*}{100-shot} & Backbone & 0.5697 & 0.5761 & 0.5678 & 0.5658 \\
    	& Fine Tuning & 0.5701 & 0.5763 & 0.5676 & 0.5669 \\
    	& Data Merging & 0.5698 & 0.5759 & 0.5697 & 0.5663 \\
    	& Reinforced Transfer Learning\citep{qu2019learning} & 0.5909 & 0.5976 & 0.5837 & 0.5885 \\
    	& Adversarial Domain Confusion\citep{cohen2018cross} & 0.5797 & 0.5857 & 0.5703 & 0.5812 \\
    	& Meta-Weight Regulator & $\textbf{0.6066}^\dagger$ & $\textbf{0.6095}^\dagger$ & $\textbf{0.6011}^\dagger$ & $\textbf{0.6003}^\dagger$ \\
    	\cline{0-5}
    	\multirow{6}{*}{1000-shot} & Backbone & 0.5854 & 0.5971 & 0.5867 & 0.5883 \\
    	& Fine Tuning & 0.5906 & 0.6051 & 0.5970 & 0.5997 \\
    	& Data Merging & 0.5835 & 0.5976 & 0.5879 & 0.5895 \\
    	& Reinforced Transfer Learning\citep{qu2019learning} & 0.6169 & 0.6206 & 0.6007 & 0.6053 \\
    	& Adversarial Domain Confusion\citep{cohen2018cross} & 0.6078 & 0.6103 & 0.6202 & \textbf{0.6205} \\
    	& Meta-Weight Regulator & $\textbf{0.6309}^\dagger$ & $\textbf{0.6397}^\dagger$ & $\textbf{0.6278}^\dagger$ & \textbf{0.6205} \\
	    \bottomrule
	\end{tabular}
	\caption{Accuracy of various adaptation methods on top of BERT (base) in the intra-task cross-dataset adaptation experiments. The experiments are conducted in four few-shot settings, where ``x-shot'' represents that each class in the target few-shot set contains x examples. The superscript $\dagger$ indicates that our method achieves statistically significant improvements over the backbone model and all other adaptation methods ($p \textless 0.05$).}
	\label{table:3}
\end{table*}

\begin{table*}[ht!]
	\centering
	\begin{tabular}{clcccccccc}  
	    \toprule
	    \multirow{2}{*}{\textbf{Target Size}} & \multirow{2}{*}{\textbf{Methods}} & \multicolumn{4}{c}{\textbf{NLI $\rightarrow$ QA}} & \multicolumn{4}{c}{\textbf{QA $\rightarrow$ NLI}} \\ \cmidrule(r){3-6} \cmidrule(r){7-10}
        & & \textbf{S} $\rightarrow$ \textbf{W} & \textbf{S} $\rightarrow$ \textbf{T} & \textbf{N} $\rightarrow$ \textbf{W} & \textbf{N} $\rightarrow$ \textbf{T} & \textbf{W} $\rightarrow$ \textbf{S} & \textbf{W} $\rightarrow$ \textbf{N} & \textbf{T} $\rightarrow$ \textbf{S} & \textbf{T} $\rightarrow$ \textbf{N}\\ 
	    \midrule
    	\multirow{6}{*}{10-shot} & Backbone & 0.5302 & 0.5323 & 0.5412 & 0.5436 & 0.5252 & 0.5238 & 0.5150 & 0.5178 \\
    	& Fine Tuning & 0.5365 & 0.5318 & 0.5465 & 0.5512 & 0.5201 & 0.5289 & 0.5189 & 0.5269 \\
    	& Data Merging & 0.5356 & 0.5281 & 0.5496 & 0.5423 & 0.5275 & 0.5285 & 0.5104 & 0.5207 \\
    	& Reinforced Transfer Learning\citep{qu2019learning} & 0.5525 & 0.5601 & 0.5652 & 0.5665 & 0.5501 & 0.5573 & 0.5592 & 0.5590 \\
    	& Adversarial Domain Confusion\citep{cohen2018cross} & 0.5521 & 0.5599 & 0.5525 & 0.5536 & 0.5395 & 0.5410 & 0.5421 & 0.5496 \\
    	& Meta-Weight Regulator & $\textbf{0.5721}^\dagger$ & \textbf{0.5710} & $\textbf{0.5855}^\dagger$ & $\textbf{0.5867}^\dagger$ & $\textbf{0.5755}^\dagger$ & \textbf{0.5732} & \textbf{0.5725} & $\textbf{0.5759}^\dagger$ \\
    	\cline{0-9}
    	\multirow{6}{*}{50-shot} & Backbone & 0.5321 & 0.5356 & 0.5499 & 0.5472 & 0.5271 & 0.5298 & 0.5199 & 0.5210 \\
    	& Fine Tuning & 0.5375 & 0.5217 & 0.5412 & 0.5524 & 0.5203 & 0.5299 & 0.5168 & 0.5277 \\
    	& Data Merging & 0.5376 & 0.5210 & 0.5510 & 0.5322 & 0.5279 & 0.5291 & 0.5201 & 0.5215 \\
    	& Reinforced Transfer Learning\citep{qu2019learning} & 0.5652 & 0.5699 & 0.5719 & 0.5705 & 0.5519 & 0.5602 & 0.5657 & 0.5612 \\
    	& Adversarial Domain Confusion\citep{cohen2018cross} & 0.5647 & 0.5691 & 0.5601 & 0.5599 & 0.5397 & 0.5413 & 0.5425 & 0.5501 \\
    	& Meta-Weight Regulator & \textbf{0.5807} & \textbf{0.5821} & $\textbf{0.5911}^\dagger$ & $\textbf{0.5904}^\dagger$ & $\textbf{0.5897}^\dagger$ & $\textbf{0.5845}^\dagger$ & \textbf{0.5823} & $\textbf{0.5929}^\dagger$ \\
    	\cline{0-9}
    	\multirow{6}{*}{100-shot} & Backbone & 0.5521 & 0.5507 & 0.5676 & 0.5689 & 0.5563 & 0.5621 & 0.5669 & 0.5687 \\
    	& Fine Tuning & 0.5537 & 0.5523 & 0.5670 & 0.5702 & 0.5615 & 0.5645 & 0.5719 & 0.5697 \\
    	& Data Merging & 0.5536 & 0.5492 & 0.5563 & 0.5613 & 0.5578 & 0.5601 & 0.5691 & 0.5539 \\
    	& Reinforced Transfer Learning\citep{qu2019learning} & 0.5755 & 0.5815 & 0.5896 & 0.5905 & 0.5776 & 0.5791 & 0.5756 & 0.5801 \\
    	& Adversarial Domain Confusion\citep{cohen2018cross} & 0.5779 & 0.5796 & 0.5834 & 0.5826 & 0.5779 & 0.5793 & 0.5757 & 0.5812 \\
    	& Meta-Weight Regulator & \textbf{0.5965} & $\textbf{0.6017}^\dagger$ & \textbf{0.5979} & $\textbf{0.6061}^\dagger$ & $\textbf{0.6003}^\dagger$ & $\textbf{0.6013}^\dagger$ & $\textbf{0.5999}^\dagger$ & $\textbf{0.6198}^\dagger$ \\
    	\cline{0-9}
    	\multirow{6}{*}{1000-shot} & Backbone & 0.5619 & 0.5625 & 0.5897 & 0.5901 & 0.5768 & 0.5724 & 0.5779 & 0.5791 \\
    	& Fine Tuning & 0.5634 & 0.5637 & 0.5870 & 0.6002 & 0.5914 & 0.5845 & 0.5896 & 0.5799 \\
    	& Data Merging & 0.5674 & 0.5729 & 0.5856 & 0.5997 & 0.5779 & 0.5809 & 0.5876 & 0.5693 \\
    	& Reinforced Transfer Learning\citep{qu2019learning} & 0.5976 & 0.5997 & 0.6103 & 0.6201 & 0.6059 & 0.6032 & 0.6075 & 0.6012 \\
    	& Adversarial Domain Confusion\citep{cohen2018cross} & 0.6077 & 0.6089 & 0.6021 & 0.6187 & \textbf{0.6125} & 0.6017 & 0.6142 & 0.6099 \\
    	& Meta-Weight Regulator & $\textbf{0.6146}^\dagger$ & $\textbf{0.6157}^\dagger$ & $\textbf{0.6275}^\dagger$ & $\textbf{0.6296}^\dagger$ & 0.6112 & $\textbf{0.6180}^\dagger$ & \textbf{0.6145} & $\textbf{0.6301}^\dagger$ \\
	    \bottomrule
	\end{tabular}
	\caption{Accuracy of various adaptation methods on top of BERT (base) in the cross-task adaptation experiments. SciTail, SNLI, WikiQA, and TrecQA are abbreviated as S, N, W, and T, respectively. The superscript $\dagger$ indicates that our method achieves statistically significant improvements over the backbone model and all other adaptation methods ($p \textless 0.05$).}
	\label{table:4}
\end{table*}

\section{Experiments}
    We empirically evaluate the proposed Meta-Weight Regulator on two typical text matching tasks, which are Question Answering and Natural Language Inference. Four widely used benchmarking datasets, i.e., SNLI \citep{bowman2015large}, SciTail \citep{khot2018scitail}, TrecQA \citep{wang2007jeopardy} and WikiQA \citep{yang2015wikiqa}, are used in the experiments on the two tasks. We classify our adaptation experiments in the few-shot learning setting into two types: \textbf{Intra-task Cross-dataset} and \textbf{Cross-task} Adaptation Experiments, which are detailed in Section 4.2.1.

    \subsection{Tasks and Datasets}
        \label{section: data}
        In our experiments, the datasets are pre-processed to ensure a consistency of class balance. The statistics for the raw and pre-processed datasets are illustrated in Table \ref{table:1} and Table \ref{table:2} respectively. The tasks and datasets used in the experiments are described in more detail as follows.
    \subsubsection{\textbf{Question Answering (QA)}}
        The goal of QA is to select the true answer for a given question from a list of candidate answers. In our experiments, we use WikiQA \citep{yang2015wikiqa} and TrecQA \citep{wang2007jeopardy}, which are  typical datasets for open domain QA.  The data of TrecQA is from the TREC Conference, while  WikiQA is constructed from Bing search queries and Wikipedia. Both data sets contain publicly available sets of question and answer pairs. Each example in the datasets is a (question, answer, label) triplet, and each label is one of {0, 1}. Nevertheless, the data distributions of the two datasets are different due to their different data sources. In these two datasets, the original numbers of positive and negative samples are not equal. They are pre-processed through downsampling to obtain a balanced number of positive and negative samples, and made consistent with the natural language inference datasets in experiments. The statistics of the processed WikiQA and TrecQA datasets are shown in the first two rows of Table \ref{table:2}.
    \subsubsection{\textbf{Natural Language Inference (NLI)}}
        NLI is a text matching task that aims to understand natural language using inference about entailment and contradiction relationships between a pair of texts. It is regarded as a valuable testing ground for the development of semantic representations. In the NLI task, we choose two widely used datasets, i.e., SNLI \citep{bowman2015large} and SciTail \citep{khot2018scitail}, which are both built via crowdsourcing. They are different in data sources. The examples of SNLI are written by humans performing a novel grounded task of describing a specific scenario from the same perspective based on image captioning. Meanwhile, SciTail is created from multiple-choice science exams and web sentences. Thus compared with SciTail, the data distribution of SNLI is more diverse. In both datasets, each example is a (premise, hypothesis, label) triplet. The labels in SNLI include "Entailment", "Neutral", and "Contradiction", while the labels in SciTail are "Entailment" and "Neutral". To keep the number of labels in the four datasets consistent in our adaptation experiments, we remove the "Contradiction" examples from SNLI. For consistency, the data in SciTail is also downsampled to achieve a class balance. The statistics for the processed SNLI and SciTail datasets are presented in the last two rows of Table \ref{table:2}.
    
    \subsection{Experimental Settings}
        \subsubsection{\textbf{The Categories of Experiments}}
            We divide the experiments into two categories according to the degree of difference between the source and target data distributions: \textbf{intra-task cross-dataset adaptation experiment} and \textbf{cross-task adaptation experiment}. In cross-task adaptation experiment, the four datasets from two tasks are cross-combined as source-target groups. For the cross-dataset adaptation experiment, we use the different datasets within a task to take turns as the source and target datasets. In each type of adaptation experiment, we compare the effectiveness of the Meta-Weight Regulator with other adaptation methods in the 10-shot, 50-shot, 100-shot, and 1000-shot settings. It may be arguable to consider 1000-shot as few-shot setting, we include it for the fairness of the experiments, to reduce the impact of class imbalance on the Adversarial Domain Confusion method (one of the baseline adaptation methods). In each set of the experiments, $T^{FS}$ is randomly sampled from the target's training set in a class-balanced manner, and the backbone model is kept consistent to ensure fair comparisons.
        \subsubsection{\textbf{Baselines}}
            We compare the proposed Meta-Weight Regulator approach, denoted as $MWR$, with the backbone models and four typical existing adaptation methods as baselines for text matching in the few-shot setting. The baseline adaptation methods are Data Merging, Fine Tuning, Adversarial Domain Confusion, and Reinforced Transfer Learning.

            \begin{itemize}
                \item \textbf{Backbone models} \citep{devlin2018bert}: We use the BERT-based-uncased as the main backbone model, for its impressive performance and capability of directly fine-tuning on train data. Here the backbone model is fine-tuned directly on $T^{FS}$ instead of being used as a fixed checkpoint. Note that we also test a range of other neural text matching models as backbones in Section \ref{section: AblationStudy} (Ablation Study). 
                \item \textbf{Data Merging (DM)}: It is a baseline adaptation method where we first merge $S^{Train}$ and $T^{FS}$ as training set, then train the backbone model on this merged training set.
                \item \textbf{Fine Tuning (FT)}: Fine Tuning method is another baseline adaptation method, where the backbone models are pre-trained on the $S^{Train}$ firstly, then they are fine-tuned on $T^{FS}$. Because the backbone model is BERT, it is fine-tuned on $S^{Train}$ and $T^{FS}$ successively.
                \item \textbf{Adversarial Domain Confusion (ADC)} \citep{cohen2018cross}: It learns a domain classifier simultaneously as model training and performs a domain confusion based on adversarial learning for cross-domain adaptation.
                \item \textbf{Reinforced Transfer Learning (RTL)} \citep{qu2019learning}: It introduces a source domain data selector based on reinforcement learning to help  transfer learning of the model to the target data.
            \end{itemize}

        \subsubsection{\textbf{Evaluation Metric}}
            For all tasks in the experiments, we adopt Cross-Entropy  as loss function and Accuracy (Acc) as the evaluation metric to measure the performance of each adaptation method on the target test set. Statistic significance is examined by permutation test with $p \textless 0.05$.

        \subsubsection{\textbf{Implementation Details}}
            In this work, the batch size of the source training set is 64, and the number of samples per class in the target few-shot set is 10, 50, 100 or 1000, making 10-shot, 50-shot, 100-shot and 1000-shot settings respectively. The hyper-parameters of text matching models are tuned on the development set, and the Meta-Weight Regulator does not bring any additional parameters. The AdamW optimizer is used to update the model parameters \citep{loshchilov2017decoupled}, and the learning rate $\alpha$ is set to $5e-5$. The maximum sequence length is truncated to 40 for QA and 50 for NLI, and padding is masked to avoid affecting the attention mechanism. For all backbone models except BERT, we use GloVe (840B tokens) to initialize the embedding look-up table for the word embedding layer, and the dimensionality of word embedding is 300. The word embedding layer is set to non-trainable. All models and methods are implemented with PyTorch \footnote{https://pytorch.org/}. The backbone BERT model we use is implemented by Huggingface \footnote{https://huggingface.co/bert-base-uncased} \citep{wolf2019huggingface}. 

    \subsection{Experimental Results}
        Here we present and analyze the results on the intra-task cross-dataset experiments and the cross-task experiments. The backbone model is jointly trained on the source training data and target few-shot data with various adaptation methods, and the final results are obtained on the target testing data.
        
        \subsubsection{\textbf{Cross-dataset Experiments}}
            Table \ref{table:3} shows the performance of the backbone model and the use of different adaptation methods on top of the backbone model for cross-dataset adaptation experiments in different few-shot settings. Although the backbone model's performance is relatively low, the first two baseline adaptation methods Fine Tuning and Data Merging, can hardly help the model improve its adaptability  and sometimes even hurt the performance. Notably, Data Merging method greatly degrades the backbone model's performance in some individual cases. We consider this may be due to the large difference between the source and target data distributions, i.e., the two distributions are far apart in the representation space. The model learns an intermediate distribution between the two on the combined data, resulting in a poor performance. The Reinforcement Transfer Learning method \citep{qu2019adversarial} benefits from the idea of data selection, which improves the performance of the backbone model  to a certain extent. It also outperforms other baseline adaptation methods when the target data in the experiments is relatively small, such as in the 10-shot, 50-shot and 100-shot settings. The Adversarial Domain Confusion method \citep{cohen2018cross} is comparable to Reinforced Transfer Learning in the 1000-shot setting. However, its performance degrades, due to its inability to meet the standard data size requirements for training classifiers as the amount of target data decreases (i.e., 10-shot, 50-shot, and 100-shot settings). 
            
            Compared with all the baseline adaptation methods, the proposed Meta-Weight Regulator achieves the best performance in almost all cases, and significantly improves the adaptability of the BERT backbone model in the few-shot settings. 

        \subsubsection{\textbf{Cross-task Experiments}}
            As shown in Table \ref{table:4}, the performance of the backbone model trained with almost every adaptation method degrades, because the difference between the source and target data distributions in cross-task experiments is larger than that in the cross-dataset experiments. Still, the Fine Tuning and  Data Merging methods  are less helpful to improve the backbone model's adaptability. Indeed, both of them even hurt the backbone model's performance in some individual cases due to the larger difference between source and target data distributions. 
            The overall performance of Reinforcement Transfer Learning \citep{qu2019adversarial} and Adversarial Domain Confusion \citep{cohen2018cross} decline compared to those in the cross-dataset experiments. Nevertheless, they still contribute to improving the backbone BERT model's adaptability. Although the overall performance of Reinforcement Transfer Learning and Adversarial Domain Confusion decline compared to those in the cross-dataset experiments, they still contribute to improving the backbone BERT model's adaptability. Unlike the results in the cross-dataset experiments, Reinforcement Transfer Learning only outperforms Adversarial Domain Confusion on fewer target data (i.e., 10-shot and 50-shot settings). However, Adversarial Domain Confusion performs on par with or better than Reinforcement Transfer Learning in the 100-shot and 1000-shot settings. This proves that simply making data selection at the data level is ineffective when the source and target data distributions are quite different. To sum up, as the data distribution gap increases, the overall performance of each method is affected to a certain extent. 

            Nonetheless, the performance of the proposed Meta-Weight Regulator still significantly surpasses all other adaptation methods in almost every set of experiments, ensuring that the overall performance does not drop considerably in comparison with the cross-dataset results. Notably, the amount of target data is roughly the same as that of source data when the source dataset is WikiQA in 1000-shot experiments, which is the ideal condition of Adversarial Domain Confusion. However, Meta-Weight Regulator still outperforms or is on par with Adversarial Domain Confusion in this setting. This result further validates the effectiveness of our approach under a higher data distribution gap.

\section{Ablation Study}
    \label{section: AblationStudy}
    In this section, we separately study the performance of Meta-Weight Regulator on other backbone text matching models, the effect of source weight initialization, and the effect of the MAML paradigm on the proposed method. The results (in average accuracy) are shown in Table \ref{table:5}, \ref{table:6}, \ref{table:7}.

\begin{table}[ht!]
	\centering
	\begin{tabular}{clcc}  
	    \toprule
	    \textbf{Backbones} & \textbf{Methods} & \textbf{Cross-Dataset} & \textbf{Cross-Task} \\ 
	    \midrule
    	\multirow{6}{*}{SiameseLSTM} & FT & 0.5382 & 0.5273 \\
    	& DM & 0.5302 & 0.5192 \\
    	& RTL\citep{qu2019learning} & 0.5522 & 0.5496 \\
    	& ADC\citep{cohen2018cross} & 0.5423 & 0.5399 \\
    	& MWR & \textbf{0.5609} & $\textbf{0.5572}^\dagger$ \\
    	\cline{0-3}
    	\multirow{6}{*}{MatchPyramid} & FT & 0.5394 & 0.5322 \\
    	& DM & 0.5366 & 0.5352 \\
    	& RTL\citep{qu2019learning} & 0.5571 & 0.5495 \\
    	& ADC\citep{cohen2018cross} & 0.5512 & 0.5436 \\
    	& MWR & $\textbf{0.5680}^\dagger$ & \textbf{0.5592} \\
    	\cline{0-3}
    	\multirow{6}{*}{Sentence-BERT} & FT & 0.5455 & 0.5336 \\
    	& DM & 0.5441 & 0.5327 \\
    	& RTL\citep{qu2019learning} & 0.5706 & 0.5645 \\
    	& ADC\citep{cohen2018cross} & 0.5571 & 0.5534 \\
    	& MWR & $\textbf{0.5870}^\dagger$ & $\textbf{0.5865}^\dagger$ \\
	    \bottomrule
	\end{tabular}
	\caption{Ablation Study on Different Backbone Models. FT, DM, RTL, ADC, and MWR stand for Fine Tuning, Data Merging, Reinforced Transfer Learning, Adversarial Domain Confusion, and Meta-Weight Regulator, respectively. The superscript $\dagger$ indicates that our method outperforms the all other methods significantly ($p \textless 0.01$).}
	\label{table:5}
\end{table}

    \subsection{Backbone Models}
        We validate the effectiveness of the Meta-Weight Regulator on three other backbone models in 50-shot cross-task and cross-dataset adaptation experiments. Note that BERT, the backbone model used in the main experiments, is an interaction-based pre-trained text matching model. The three additional backbone models represent a variety of alternatives, including the interaction-based text matching model MatchPyramid, the representation-based model SiameseLSTM, and another pre-trained model Sentence-BERT. 
        Furthermore, MatchPyramid and SiameseLSTM are not pre-trained models like BERT and Sentence-BERT, so that they cannot be directly fine-tuned on $T^{FS}$. Accordingly the Fine Tuning (FT) method for them in experiments refers to pre-training them on $S^{Train}$ and then fine-tuning on $T^{FS}$.
        Table \ref{table:5} demonstrates a better performance of the proposed Meta-Weight Regulator than other adaptation methods, when applied to the different backbone text matching models. This result proves that the Meta-Weight Regulator does not depend on the adequate performance of BERT, and shows that our approach can improve adaptability for a wide range of text matching models.

\begin{table}
	\centering
	\begin{tabular}{l|cc}  
	\toprule
    \textbf{Source Weight} & \textbf{Cross-Dataset} & \textbf{Cross-Task} \\
	\midrule
    	Random Initialization & 0.5425 & 0.5376 \\
    	One Initialization & 0.5796 & 0.5751 \\
    	Zero Initialization & \textbf{0.5871} & \textbf{0.5867} \\
	\bottomrule
	\end{tabular}
	\caption{Ablation Study on Source Weight Initialization.}
	\label{table:6}
\end{table}

\begin{table}
	\centering
	\begin{tabular}{c|cc}  
	\toprule
    \textbf{Meta-learning} & \textbf{Cross-Dataset} & \textbf{Cross-Task} \\
	\midrule
    	Reptile & 0.5828 & 0.5809 \\
    	MAML & \textbf{0.5871} & \textbf{0.5867} \\ 
	\bottomrule
	\end{tabular}
	\caption{Ablation Study on Meta-Learning Methods.}
	\label{table:7}
\end{table}

    \subsection{Source Weight Initialization}
       In this part, we ablate the effect of source weight initialization by contrasting the zero initialization with the other initialization on the 50-shot adaptation experiments. One and Random Initialization for weight are used to compare Zero Initialization in Meta-Weight Regulator. We use BERT-base-uncased as the backbone model and train it with above three variants of Meta-Weight Regulator on adaptation experiments. The performance is shown in Table \ref{table:6}, which demonstrates that:
        \begin{enumerate}
            \item The performance of random initialization is poor. This is probably because the number of examples in the target few-shot set is insufficient to update the overly scattered initialization values to appropriate values.
            \item Zero Initialization for source weight is better than the One Initialization. We consider that this is due to One Initialization assigning source samples too much weights that are equal to that of target samples, so that the models can not distinguish which of them are more important.
            \item Zero initialization for source weights is preferable to the random initialization. The root cause may be that random initialization can easily trap model parameters into local minima.
        \end{enumerate}

    \subsection{Effectiveness of MAML}
    At the end of this section, we discuss the meta-learning paradigm type for Meta-Weight Regulator. Similarly, we perform ablation experiments based on BERT on 50-shot adaptation experiments to verify the effectiveness of MAML \citep{finn2017model}. As shown in Table \ref{table:7}, the regulator using MAML is slightly better than Reptile \citep{nichol2018reptile}, and we make the following analysis:
    \begin{enumerate}
        \item Unlike MAML’s meta-gradient descent, which is only updated once, Reptile can update multiple times on each data distribution when training the network.
        \item To improve the engineering efficiency, Reptile does not calculate high-order derivatives to update like MAML but directly introduces a parameter multiplied by the difference between the meta-network and the trained network parameters to update the parameters of the meta-network.
        \item Although the regulator using MAML does not outperform its counterpart using Reptile by a large margin, MAML does not introduce additional hyperparameters, thus improving training efficiency.
    \end{enumerate}

\section{Conclusions and Future Work}
    In this paper, we have proposed a Meta-Weight Regulator for cross-dataset and cross-task adaptation of deep text matching models in the few-shot setting. Specifically, we use the meta-learning paradigm to regulate the weights of source training examples in each batch according to the target few-shot loss. Our method imposes no constraints on the data and can be trained on source and target data simultaneously for gradient-level data selection. We construct intra-task cross-dataset and cross-task adaptation experiments using four benchmark datasets over two text matching tasks (question answering and natural language perturbation) in different few-shot settings. Extensive experiments demonstrate that the Meta-Weight Regulator can significantly outperform a range of state-of-the-art adaptation methods regardless of how different the source and target data distributions are. Furthermore, our method effectively enhances the adaptability of the text matching models and does not introduce any extra hyper-parameters. In the future, we would like to extend our approach to generalize deep text matching models to the target distribution without any target data. 

\section{Acknowledgements}
    This research was supported in part by Huawei Technologies (project number: TC20201228005). 

\bibliographystyle{ACM-Reference-Format}
\balance
\bibliography{references.bib}


\begin{thebibliography}{37}


\ifx \showCODEN    \undefined \def \showCODEN     #1{\unskip}     \fi
\ifx \showDOI      \undefined \def \showDOI       #1{#1}\fi
\ifx \showISBNx    \undefined \def \showISBNx     #1{\unskip}     \fi
\ifx \showISBNxiii \undefined \def \showISBNxiii  #1{\unskip}     \fi
\ifx \showISSN     \undefined \def \showISSN      #1{\unskip}     \fi
\ifx \showLCCN     \undefined \def \showLCCN      #1{\unskip}     \fi
\ifx \shownote     \undefined \def \shownote      #1{#1}          \fi
\ifx \showarticletitle \undefined \def \showarticletitle #1{#1}   \fi
\ifx \showURL      \undefined \def \showURL       {\relax}        \fi
\providecommand\bibfield[2]{#2}
\providecommand\bibinfo[2]{#2}
\providecommand\natexlab[1]{#1}
\providecommand\showeprint[2][]{arXiv:#2}

\bibitem[Asghar et~al\mbox{.}(2018)]%
        {asghar2018progressive}
\bibfield{author}{\bibinfo{person}{Nabiha Asghar}, \bibinfo{person}{Lili Mou},
  \bibinfo{person}{Kira~A Selby}, \bibinfo{person}{Kevin~D Pantasdo},
  \bibinfo{person}{Pascal Poupart}, {and} \bibinfo{person}{Xin Jiang}.}
  \bibinfo{year}{2018}\natexlab{}.
\newblock \showarticletitle{Progressive memory banks for incremental domain
  adaptation}.
\newblock \bibinfo{journal}{\emph{arXiv preprint arXiv:1811.00239}}
  (\bibinfo{year}{2018}).
\newblock


\bibitem[Balaji et~al\mbox{.}(2020)]%
        {balaji2020robust}
\bibfield{author}{\bibinfo{person}{Yogesh Balaji}, \bibinfo{person}{Rama
  Chellappa}, {and} \bibinfo{person}{Soheil Feizi}.}
  \bibinfo{year}{2020}\natexlab{}.
\newblock \showarticletitle{Robust optimal transport with applications in
  generative modeling and domain adaptation}.
\newblock \bibinfo{journal}{\emph{arXiv preprint arXiv:2010.05862}}
  (\bibinfo{year}{2020}).
\newblock


\bibitem[Bengio et~al\mbox{.}(2009)]%
        {bengio2009curriculum}
\bibfield{author}{\bibinfo{person}{Yoshua Bengio},
  \bibinfo{person}{J{\'e}r{\^o}me Louradour}, \bibinfo{person}{Ronan
  Collobert}, {and} \bibinfo{person}{Jason Weston}.}
  \bibinfo{year}{2009}\natexlab{}.
\newblock \showarticletitle{Curriculum learning}. In
  \bibinfo{booktitle}{\emph{Proceedings of the 26th annual international
  conference on machine learning}}. \bibinfo{pages}{41--48}.
\newblock


\bibitem[Bowman et~al\mbox{.}(2015)]%
        {bowman2015large}
\bibfield{author}{\bibinfo{person}{Samuel~R Bowman}, \bibinfo{person}{Gabor
  Angeli}, \bibinfo{person}{Christopher Potts}, {and}
  \bibinfo{person}{Christopher~D Manning}.} \bibinfo{year}{2015}\natexlab{}.
\newblock \showarticletitle{A large annotated corpus for learning natural
  language inference}.
\newblock \bibinfo{journal}{\emph{arXiv preprint arXiv:1508.05326}}
  (\bibinfo{year}{2015}).
\newblock


\bibitem[Chen et~al\mbox{.}(2016)]%
        {chen2016enhanced}
\bibfield{author}{\bibinfo{person}{Qian Chen}, \bibinfo{person}{Xiaodan Zhu},
  \bibinfo{person}{Zhenhua Ling}, \bibinfo{person}{Si Wei},
  \bibinfo{person}{Hui Jiang}, {and} \bibinfo{person}{Diana Inkpen}.}
  \bibinfo{year}{2016}\natexlab{}.
\newblock \showarticletitle{Enhanced lstm for natural language inference}.
\newblock \bibinfo{journal}{\emph{arXiv preprint arXiv:1609.06038}}
  (\bibinfo{year}{2016}).
\newblock


\bibitem[Cherry et~al\mbox{.}(1998)]%
        {cherry1998sgd}
\bibfield{author}{\bibinfo{person}{J~Michael Cherry}, \bibinfo{person}{Caroline
  Adler}, \bibinfo{person}{Catherine Ball}, \bibinfo{person}{Stephen~A
  Chervitz}, \bibinfo{person}{Selina~S Dwight}, \bibinfo{person}{Erich~T
  Hester}, \bibinfo{person}{Yankai Jia}, \bibinfo{person}{Gail Juvik},
  \bibinfo{person}{TaiYun Roe}, \bibinfo{person}{Mark Schroeder},
  {et~al\mbox{.}}} \bibinfo{year}{1998}\natexlab{}.
\newblock \showarticletitle{SGD: Saccharomyces genome database}.
\newblock \bibinfo{journal}{\emph{Nucleic acids research}}
  \bibinfo{volume}{26}, \bibinfo{number}{1} (\bibinfo{year}{1998}),
  \bibinfo{pages}{73--79}.
\newblock


\bibitem[Cohen et~al\mbox{.}(2018)]%
        {cohen2018cross}
\bibfield{author}{\bibinfo{person}{Daniel Cohen}, \bibinfo{person}{Bhaskar
  Mitra}, \bibinfo{person}{Katja Hofmann}, {and} \bibinfo{person}{W~Bruce
  Croft}.} \bibinfo{year}{2018}\natexlab{}.
\newblock \showarticletitle{Cross domain regularization for neural ranking
  models using adversarial learning}. In \bibinfo{booktitle}{\emph{The 41st
  International ACM SIGIR Conference on Research \& Development in Information
  Retrieval}}. \bibinfo{pages}{1025--1028}.
\newblock


\bibitem[Devlin et~al\mbox{.}(2018)]%
        {devlin2018bert}
\bibfield{author}{\bibinfo{person}{Jacob Devlin}, \bibinfo{person}{Ming-Wei
  Chang}, \bibinfo{person}{Kenton Lee}, {and} \bibinfo{person}{Kristina
  Toutanova}.} \bibinfo{year}{2018}\natexlab{}.
\newblock \showarticletitle{Bert: Pre-training of deep bidirectional
  transformers for language understanding}.
\newblock \bibinfo{journal}{\emph{arXiv preprint arXiv:1810.04805}}
  (\bibinfo{year}{2018}).
\newblock


\bibitem[Feng et~al\mbox{.}(2015)]%
        {feng2015applying}
\bibfield{author}{\bibinfo{person}{Minwei Feng}, \bibinfo{person}{Bing Xiang},
  \bibinfo{person}{Michael~R Glass}, \bibinfo{person}{Lidan Wang}, {and}
  \bibinfo{person}{Bowen Zhou}.} \bibinfo{year}{2015}\natexlab{}.
\newblock \showarticletitle{Applying deep learning to answer selection: A study
  and an open task}. In \bibinfo{booktitle}{\emph{2015 IEEE Workshop on
  Automatic Speech Recognition and Understanding (ASRU)}}. IEEE,
  \bibinfo{pages}{813--820}.
\newblock


\bibitem[Finn et~al\mbox{.}(2017)]%
        {finn2017model}
\bibfield{author}{\bibinfo{person}{Chelsea Finn}, \bibinfo{person}{Pieter
  Abbeel}, {and} \bibinfo{person}{Sergey Levine}.}
  \bibinfo{year}{2017}\natexlab{}.
\newblock \showarticletitle{Model-agnostic meta-learning for fast adaptation of
  deep networks}. In \bibinfo{booktitle}{\emph{International Conference on
  Machine Learning}}. PMLR, \bibinfo{pages}{1126--1135}.
\newblock


\bibitem[Freund and Schapire(1997)]%
        {freund1997decision}
\bibfield{author}{\bibinfo{person}{Yoav Freund} {and} \bibinfo{person}{Robert~E
  Schapire}.} \bibinfo{year}{1997}\natexlab{}.
\newblock \showarticletitle{A decision-theoretic generalization of on-line
  learning and an application to boosting}.
\newblock \bibinfo{journal}{\emph{Journal of computer and system sciences}}
  \bibinfo{volume}{55}, \bibinfo{number}{1} (\bibinfo{year}{1997}),
  \bibinfo{pages}{119--139}.
\newblock


\bibitem[Ganin et~al\mbox{.}(2016)]%
        {ganin2016domain}
\bibfield{author}{\bibinfo{person}{Yaroslav Ganin}, \bibinfo{person}{Evgeniya
  Ustinova}, \bibinfo{person}{Hana Ajakan}, \bibinfo{person}{Pascal Germain},
  \bibinfo{person}{Hugo Larochelle}, \bibinfo{person}{Fran{\c{c}}ois
  Laviolette}, \bibinfo{person}{Mario Marchand}, {and} \bibinfo{person}{Victor
  Lempitsky}.} \bibinfo{year}{2016}\natexlab{}.
\newblock \showarticletitle{Domain-adversarial training of neural networks}.
\newblock \bibinfo{journal}{\emph{The journal of machine learning research}}
  \bibinfo{volume}{17}, \bibinfo{number}{1} (\bibinfo{year}{2016}),
  \bibinfo{pages}{2096--2030}.
\newblock


\bibitem[Jiang et~al\mbox{.}(2017)]%
        {jiangmentornet}
\bibfield{author}{\bibinfo{person}{L Jiang}, \bibinfo{person}{Z Zhou},
  \bibinfo{person}{T Leung}, \bibinfo{person}{L Li}, {and} \bibinfo{person}{L
  Fei-Fei}.} \bibinfo{year}{2017}\natexlab{}.
\newblock \bibinfo{title}{Mentornet: Regularizing very deep neural networks on
  corrupted labels. CoRR abs/1712.05055 (2017)}.
\newblock
\newblock


\bibitem[Kahn and Marshall(1953)]%
        {kahn1953methods}
\bibfield{author}{\bibinfo{person}{Herman Kahn} {and} \bibinfo{person}{Andy~W
  Marshall}.} \bibinfo{year}{1953}\natexlab{}.
\newblock \showarticletitle{Methods of reducing sample size in Monte Carlo
  computations}.
\newblock \bibinfo{journal}{\emph{Journal of the Operations Research Society of
  America}} \bibinfo{volume}{1}, \bibinfo{number}{5} (\bibinfo{year}{1953}),
  \bibinfo{pages}{263--278}.
\newblock


\bibitem[Khot et~al\mbox{.}(2018)]%
        {khot2018scitail}
\bibfield{author}{\bibinfo{person}{Tushar Khot}, \bibinfo{person}{Ashish
  Sabharwal}, {and} \bibinfo{person}{Peter Clark}.}
  \bibinfo{year}{2018}\natexlab{}.
\newblock \showarticletitle{Scitail: A textual entailment dataset from science
  question answering}. In \bibinfo{booktitle}{\emph{Thirty-Second AAAI
  Conference on Artificial Intelligence}}.
\newblock


\bibitem[Kingma and Ba(2014)]%
        {kingma2014adam}
\bibfield{author}{\bibinfo{person}{Diederik~P Kingma} {and}
  \bibinfo{person}{Jimmy Ba}.} \bibinfo{year}{2014}\natexlab{}.
\newblock \showarticletitle{Adam: A method for stochastic optimization}.
\newblock \bibinfo{journal}{\emph{arXiv preprint arXiv:1412.6980}}
  (\bibinfo{year}{2014}).
\newblock


\bibitem[Koehn and Knowles(2017)]%
        {koehn2017six}
\bibfield{author}{\bibinfo{person}{Philipp Koehn} {and}
  \bibinfo{person}{Rebecca Knowles}.} \bibinfo{year}{2017}\natexlab{}.
\newblock \showarticletitle{Six challenges for neural machine translation}.
\newblock \bibinfo{journal}{\emph{arXiv preprint arXiv:1706.03872}}
  (\bibinfo{year}{2017}).
\newblock


\bibitem[Kumar et~al\mbox{.}(2010)]%
        {kumar2010self}
\bibfield{author}{\bibinfo{person}{M~Pawan Kumar}, \bibinfo{person}{Benjamin
  Packer}, {and} \bibinfo{person}{Daphne Koller}.}
  \bibinfo{year}{2010}\natexlab{}.
\newblock \showarticletitle{Self-Paced Learning for Latent Variable Models.}.
  In \bibinfo{booktitle}{\emph{NIPS}}, Vol.~\bibinfo{volume}{1}.
  \bibinfo{pages}{2}.
\newblock


\bibitem[Li et~al\mbox{.}(2016)]%
        {li2016revisiting}
\bibfield{author}{\bibinfo{person}{Yanghao Li}, \bibinfo{person}{Naiyan Wang},
  \bibinfo{person}{Jianping Shi}, \bibinfo{person}{Jiaying Liu}, {and}
  \bibinfo{person}{Xiaodi Hou}.} \bibinfo{year}{2016}\natexlab{}.
\newblock \showarticletitle{Revisiting batch normalization for practical domain
  adaptation}.
\newblock \bibinfo{journal}{\emph{arXiv preprint arXiv:1603.04779}}
  (\bibinfo{year}{2016}).
\newblock


\bibitem[Lin et~al\mbox{.}(2017)]%
        {lin2017focal}
\bibfield{author}{\bibinfo{person}{Tsung-Yi Lin}, \bibinfo{person}{Priya
  Goyal}, \bibinfo{person}{Ross Girshick}, \bibinfo{person}{Kaiming He}, {and}
  \bibinfo{person}{Piotr Doll{\'a}r}.} \bibinfo{year}{2017}\natexlab{}.
\newblock \showarticletitle{Focal loss for dense object detection}. In
  \bibinfo{booktitle}{\emph{Proceedings of the IEEE international conference on
  computer vision}}. \bibinfo{pages}{2980--2988}.
\newblock


\bibitem[Loshchilov and Hutter(2017)]%
        {loshchilov2017decoupled}
\bibfield{author}{\bibinfo{person}{Ilya Loshchilov} {and}
  \bibinfo{person}{Frank Hutter}.} \bibinfo{year}{2017}\natexlab{}.
\newblock \showarticletitle{Decoupled weight decay regularization}.
\newblock \bibinfo{journal}{\emph{arXiv preprint arXiv:1711.05101}}
  (\bibinfo{year}{2017}).
\newblock


\bibitem[Malisiewicz et~al\mbox{.}(2011)]%
        {malisiewicz2011ensemble}
\bibfield{author}{\bibinfo{person}{Tomasz Malisiewicz},
  \bibinfo{person}{Abhinav Gupta}, {and} \bibinfo{person}{Alexei~A Efros}.}
  \bibinfo{year}{2011}\natexlab{}.
\newblock \showarticletitle{Ensemble of exemplar-svms for object detection and
  beyond}. In \bibinfo{booktitle}{\emph{2011 International conference on
  computer vision}}. IEEE, \bibinfo{pages}{89--96}.
\newblock


\bibitem[Mueller and Thyagarajan(2016)]%
        {mueller2016siamese}
\bibfield{author}{\bibinfo{person}{Jonas Mueller} {and} \bibinfo{person}{Aditya
  Thyagarajan}.} \bibinfo{year}{2016}\natexlab{}.
\newblock \showarticletitle{Siamese recurrent architectures for learning
  sentence similarity}. In \bibinfo{booktitle}{\emph{Proceedings of the AAAI
  conference on artificial intelligence}}, Vol.~\bibinfo{volume}{30}.
\newblock


\bibitem[Nichol and Schulman(2018)]%
        {nichol2018reptile}
\bibfield{author}{\bibinfo{person}{Alex Nichol} {and} \bibinfo{person}{John
  Schulman}.} \bibinfo{year}{2018}\natexlab{}.
\newblock \showarticletitle{Reptile: a scalable metalearning algorithm}.
\newblock \bibinfo{journal}{\emph{arXiv preprint arXiv:1803.02999}}
  \bibinfo{volume}{2}, \bibinfo{number}{3} (\bibinfo{year}{2018}),
  \bibinfo{pages}{4}.
\newblock


\bibitem[Nie and Bansal(2017)]%
        {nie2017shortcut}
\bibfield{author}{\bibinfo{person}{Yixin Nie} {and} \bibinfo{person}{Mohit
  Bansal}.} \bibinfo{year}{2017}\natexlab{}.
\newblock \showarticletitle{Shortcut-stacked sentence encoders for multi-domain
  inference}.
\newblock \bibinfo{journal}{\emph{arXiv preprint arXiv:1708.02312}}
  (\bibinfo{year}{2017}).
\newblock


\bibitem[Pang et~al\mbox{.}(2016)]%
        {pang2016text}
\bibfield{author}{\bibinfo{person}{Liang Pang}, \bibinfo{person}{Yanyan Lan},
  \bibinfo{person}{Jiafeng Guo}, \bibinfo{person}{Jun Xu},
  \bibinfo{person}{Shengxian Wan}, {and} \bibinfo{person}{Xueqi Cheng}.}
  \bibinfo{year}{2016}\natexlab{}.
\newblock \showarticletitle{Text matching as image recognition}. In
  \bibinfo{booktitle}{\emph{Proceedings of the AAAI Conference on Artificial
  Intelligence}}, Vol.~\bibinfo{volume}{30}.
\newblock


\bibitem[Qu et~al\mbox{.}(2019a)]%
        {qu2019learning}
\bibfield{author}{\bibinfo{person}{Chen Qu}, \bibinfo{person}{Feng Ji},
  \bibinfo{person}{Minghui Qiu}, \bibinfo{person}{Liu Yang},
  \bibinfo{person}{Zhiyu Min}, \bibinfo{person}{Haiqing Chen},
  \bibinfo{person}{Jun Huang}, {and} \bibinfo{person}{W~Bruce Croft}.}
  \bibinfo{year}{2019}\natexlab{a}.
\newblock \showarticletitle{Learning to selectively transfer: Reinforced
  transfer learning for deep text matching}. In
  \bibinfo{booktitle}{\emph{Proceedings of the Twelfth ACM International
  Conference on Web Search and Data Mining}}. \bibinfo{pages}{699--707}.
\newblock


\bibitem[Qu et~al\mbox{.}(2019b)]%
        {qu2019adversarial}
\bibfield{author}{\bibinfo{person}{Xiaoye Qu}, \bibinfo{person}{Zhikang Zou},
  \bibinfo{person}{Yu Cheng}, \bibinfo{person}{Yang Yang}, {and}
  \bibinfo{person}{Pan Zhou}.} \bibinfo{year}{2019}\natexlab{b}.
\newblock \showarticletitle{Adversarial category alignment network for
  cross-domain sentiment classification}. In
  \bibinfo{booktitle}{\emph{Proceedings of the 2019 Conference of the North
  American Chapter of the Association for Computational Linguistics: Human
  Language Technologies, Volume 1 (Long and Short Papers)}}.
  \bibinfo{pages}{2496--2508}.
\newblock


\bibitem[Reimers and Gurevych(2019)]%
        {reimers2019sentence}
\bibfield{author}{\bibinfo{person}{Nils Reimers} {and} \bibinfo{person}{Iryna
  Gurevych}.} \bibinfo{year}{2019}\natexlab{}.
\newblock \showarticletitle{Sentence-bert: Sentence embeddings using siamese
  bert-networks}.
\newblock \bibinfo{journal}{\emph{arXiv preprint arXiv:1908.10084}}
  (\bibinfo{year}{2019}).
\newblock


\bibitem[Ren et~al\mbox{.}(2018)]%
        {ren2018learning}
\bibfield{author}{\bibinfo{person}{Mengye Ren}, \bibinfo{person}{Wenyuan Zeng},
  \bibinfo{person}{Bin Yang}, {and} \bibinfo{person}{Raquel Urtasun}.}
  \bibinfo{year}{2018}\natexlab{}.
\newblock \showarticletitle{Learning to reweight examples for robust deep
  learning}. In \bibinfo{booktitle}{\emph{International Conference on Machine
  Learning}}. PMLR, \bibinfo{pages}{4334--4343}.
\newblock


\bibitem[Wang et~al\mbox{.}(2007)]%
        {wang2007jeopardy}
\bibfield{author}{\bibinfo{person}{Mengqiu Wang}, \bibinfo{person}{Noah~A
  Smith}, {and} \bibinfo{person}{Teruko Mitamura}.}
  \bibinfo{year}{2007}\natexlab{}.
\newblock \showarticletitle{What is the Jeopardy model? A quasi-synchronous
  grammar for QA}. In \bibinfo{booktitle}{\emph{Proceedings of the 2007 Joint
  Conference on Empirical Methods in Natural Language Processing and
  Computational Natural Language Learning (EMNLP-CoNLL)}}.
  \bibinfo{pages}{22--32}.
\newblock


\bibitem[Wolf et~al\mbox{.}(2019)]%
        {wolf2019huggingface}
\bibfield{author}{\bibinfo{person}{Thomas Wolf}, \bibinfo{person}{Lysandre
  Debut}, \bibinfo{person}{Victor Sanh}, \bibinfo{person}{Julien Chaumond},
  \bibinfo{person}{Clement Delangue}, \bibinfo{person}{Anthony Moi},
  \bibinfo{person}{Pierric Cistac}, \bibinfo{person}{Tim Rault},
  \bibinfo{person}{R{\'e}mi Louf}, \bibinfo{person}{Morgan Funtowicz},
  {et~al\mbox{.}}} \bibinfo{year}{2019}\natexlab{}.
\newblock \showarticletitle{Huggingface's transformers: State-of-the-art
  natural language processing}.
\newblock \bibinfo{journal}{\emph{arXiv preprint arXiv:1910.03771}}
  (\bibinfo{year}{2019}).
\newblock


\bibitem[Xue et~al\mbox{.}(2020)]%
        {xue2020improving}
\bibfield{author}{\bibinfo{person}{Qianming Xue}, \bibinfo{person}{Wei Zhang},
  {and} \bibinfo{person}{Hongyuan Zha}.} \bibinfo{year}{2020}\natexlab{}.
\newblock \showarticletitle{Improving domain-adapted sentiment classification
  by deep adversarial mutual learning}. In
  \bibinfo{booktitle}{\emph{Proceedings of the AAAI Conference on Artificial
  Intelligence}}, Vol.~\bibinfo{volume}{34}. \bibinfo{pages}{9362--9369}.
\newblock


\bibitem[Yang and Soatto(2020)]%
        {yang2020fda}
\bibfield{author}{\bibinfo{person}{Yanchao Yang} {and} \bibinfo{person}{Stefano
  Soatto}.} \bibinfo{year}{2020}\natexlab{}.
\newblock \showarticletitle{Fda: Fourier domain adaptation for semantic
  segmentation}. In \bibinfo{booktitle}{\emph{Proceedings of the IEEE/CVF
  Conference on Computer Vision and Pattern Recognition}}.
  \bibinfo{pages}{4085--4095}.
\newblock


\bibitem[Yang et~al\mbox{.}(2015)]%
        {yang2015wikiqa}
\bibfield{author}{\bibinfo{person}{Yi Yang}, \bibinfo{person}{Wen-tau Yih},
  {and} \bibinfo{person}{Christopher Meek}.} \bibinfo{year}{2015}\natexlab{}.
\newblock \showarticletitle{Wikiqa: A challenge dataset for open-domain
  question answering}. In \bibinfo{booktitle}{\emph{Proceedings of the 2015
  conference on empirical methods in natural language processing}}.
  \bibinfo{pages}{2013--2018}.
\newblock


\bibitem[Zhang et~al\mbox{.}(2020)]%
        {zhang2020domain}
\bibfield{author}{\bibinfo{person}{Kun Zhang}, \bibinfo{person}{Mingming Gong},
  \bibinfo{person}{Petar Stojanov}, \bibinfo{person}{Biwei Huang},
  \bibinfo{person}{Qingsong Liu}, {and} \bibinfo{person}{Clark Glymour}.}
  \bibinfo{year}{2020}\natexlab{}.
\newblock \showarticletitle{Domain adaptation as a problem of inference on
  graphical models}.
\newblock \bibinfo{journal}{\emph{arXiv preprint arXiv:2002.03278}}
  (\bibinfo{year}{2020}).
\newblock


\bibitem[Zhao et~al\mbox{.}(2020)]%
        {zhao2020domain}
\bibfield{author}{\bibinfo{person}{Shanshan Zhao}, \bibinfo{person}{Mingming
  Gong}, \bibinfo{person}{Tongliang Liu}, \bibinfo{person}{Huan Fu}, {and}
  \bibinfo{person}{Dacheng Tao}.} \bibinfo{year}{2020}\natexlab{}.
\newblock \showarticletitle{Domain generalization via entropy regularization}.
\newblock \bibinfo{journal}{\emph{Advances in Neural Information Processing
  Systems}}  \bibinfo{volume}{33} (\bibinfo{year}{2020}).
\newblock


\end{thebibliography}

\end{document}